%% file: arxiv_submission.tex
\documentclass{IEEEtran4PSCC}

\usepackage{graphicx}
\graphicspath{ {./images/} }
\usepackage[cmex10]{amsmath}
\usepackage{array}
\usepackage{amssymb}
\usepackage{algorithm}
\usepackage{algpseudocode} 
\usepackage{subcaption}
\input{defs}

\usepackage[style=ieee, backend=biber]{biblatex}
\usepackage{multirow}
\usepackage{tikz}
\usetikzlibrary{%
  arrows,
  positioning
}
\usetikzlibrary{intersections}

\usepackage{tikz-qtree}
\usetikzlibrary{positioning,shapes.multipart}

\usepackage{pgfplots}
\usetikzlibrary{fillbetween}

\renewcommand{\baselinestretch}{0.993}

\allowdisplaybreaks

\makeatletter
\let\old@ps@headings\ps@headings
\let\old@ps@IEEEtitlepagestyle\ps@IEEEtitlepagestyle
\def\psccfooter#1{%
    \def\ps@headings{%
        \old@ps@headings%
        \def\@oddfoot{\strut\hfill#1\hfill\strut}%
        \def\@evenfoot{\strut\hfill#1\hfill\strut}%
    }%
    \def\ps@IEEEtitlepagestyle{%
        \old@ps@IEEEtitlepagestyle%
        \def\@oddfoot{\strut\hfill#1\hfill\strut}%
        \def\@evenfoot{\strut\hfill#1\hfill\strut}%
    }%
    \ps@headings%
}
\makeatother

\begin{document}
%
\title{Optimally Managing the Impacts of Convergence Tolerance for Distributed Optimal Power Flow}

 \author{\IEEEauthorblockN{Rachel Harris, Mohannad Alkhraijah,
and
 Daniel K. Molzahn}
 }

\maketitle

\begin{abstract}
The future power grid may rely on distributed optimization to determine the set-points for huge numbers of distributed energy resources. There has been significant work on applying distributed algorithms to optimal power flow (OPF) problems, which require separate computing agents to agree on shared boundary variable values. 
Looser tolerances for the mismatches in these shared variables generally yield faster convergence at the expense of exacerbating constraint violations, but there is little quantitative understanding of how the convergence tolerance affects solution quality. 
To address this gap, we first quantify how convergence tolerance impacts constraint violations when the distributed OPF generator dispatch is applied to the power system. Using insights from this analysis, we then develop a bound tightening algorithm which guarantees that operating points from distributed OPF algorithms will not result in violations despite the possibility of shared variable mismatches within the convergence tolerance. We also explore how bounding the cumulative shared variable mismatches can prevent unnecessary conservativeness in the bound tightening. The proposed approach enables control of the trade-off between computational speed, which improves as the convergence tolerance increases, and distributed OPF solution cost, which increases with convergence tolerance due to tightened constraints, while ensuring feasibility.%
\end{abstract}%


\thanksto{\noindent The authors are with the School of Electrical and Computer Engineering, Georgia Institute of Technology, Atlanta, GA, USA. Support from NSF AI Institute for Advances in Optimization (AI4OPT), \#2112533.}

\section{Introduction}
As we transition to low-carbon power systems, distributed energy resources (DERs) such as electric vehicles, battery storage systems, and wind and solar generators will increase by orders of magnitude, motivating the development of new optimization and control methods~\cite{molzahn2017survey}. Traditional power system optimization approaches where a central operator collects system-wide information and computes optimal dispatches for bulk generation plants may be inadequate for future power systems with widespread DER integration and consumers who desire data privacy. Distributed optimization algorithms can scale to large, complex problems, have the potential to keep local information private, and avoid a single point of failure. 

Many researchers have applied distributed algorithms to the optimal power flow (OPF) problem. Commonly used distributed algorithms include the alternating direction method of multipliers (ADMM) \cite{6748974,6672864}, analytical target cascading (ATC) \cite{atc,6780985}, and auxiliary problem principle (APP) \cite{app_convergence,app2}. Distributed algorithms decompose the system into separate regions, each under the control of different local computing agents. These agents solve local optimization problems and share boundary variable values to ensure consistency between regions. The algorithm converges when the norm of the shared variable mismatch values falls below a convergence tolerance $\epsilon$. The authors of \cite{admm_book} provide some guidance on how to select $\epsilon$ based on scale of the variables, and most researchers select $\epsilon \in [10^{-5},10^{-3}]$. However, to the best of our knowledge, the literature contains no detailed analysis of the impact of convergence tolerance on constraint violations after a distributed OPF solution is applied to the power grid. 

Many distributed algorithms take thousands of iterations to converge for large-scale systems~\cite{7776940,WANG2017127,7581042}. To use such distributed algorithms to operate future power grids with many rapidly fluctuating DERs, we must reduce distributed OPF computation time. To accelerate distributed algorithms, some researchers have proposed adaptive parameter tuning~\cite{8573893,8994051,zeng_kody_kim_kim_molzahn-rl_for_dist_opt} and using machine learning to predict the converged boundary variable values~\cite{9482738,10057067}. One simple way to reduce convergence time is to select a larger convergence tolerance $\epsilon$. As we will demonstrate in this paper, looser tolerances can significantly decrease the the number of iterations required to converge and thus reduce computation times. However, before loosening the tolerance, we must ensure the resulting distributed OPF solution provides a safe operating point that will not cause constraint violations.  

In this paper, we assess the impacts of convergence tolerance on constraint violations and develop a bound tightening algorithm which prevents these violations. We focus on the AC OPF problem solved with the ADMM distributed algorithm, but our method can be applied without any conceptual changes to other power flow formulations or distributed algorithms. We formulate an optimization problem which, given some convergence tolerance, finds the maximum possible violation for each constraint at the power system operating point under distributed OPF dispatch. Next, we propose an algorithm which iterates between finding these maximum violations and updating bound tightenings to guarantee that distributed OPF with these tightened bounds will not result in any constraint violations. We present numerical results from several representative test cases, showing that running distributed OPF on the bound-tightened cases significantly decreases computation time without resulting in constraint violations. 

The remainder of this paper is organized as follows. In Section~\ref{sec:problem_formulation}, we describe the distributed OPF formulation and discuss how the choice of convergence tolerance impacts computation speed as well as feasibility and optimality of the final solution. Section~\ref{sec:analysis_and_algorithms} formulates an optimization problem which finds the worst-case constraint violations that may result from selecting a certain convergence tolerance for the distributed OPF computation. We also present a bound tightening algorithm which iteratively solves this optimization problem and tightens constraints until there can be no violations of the original constraints when the distributed OPF solution with the given convergence tolerance is applied to the system. The algorithm may be augmented with bounds on cumulative mismatches so that bound tightening is less conservative. In Section~\ref{sec:results}, we present numerical results, including solution costs for constraint-tightened test cases and relationships between cumulative mismatch bounds and violations. We conclude and discuss future work in Section~\ref{sec:conclusion}. 

\section{Distributed OPF Formulation}\label{sec:problem_formulation}
This section provides background material by formulating the OPF problem and reviewing distributed OPF algorithms.


\subsection{Optimal Power Flow}\label{sec:opf}
The OPF problem optimizes performance subject to operational limits and physical power flow equations. In this paper, we consider OPF formulations with an AC power flow model and an objective that minimizes generation cost as a quadratic function of real power output from each generator. However, alternative power flow formulations or different objectives could also be used. 

Let $\mathcal{N}$, $\mathcal{E}$, and $\mathcal{G}$ denote the sets of buses, lines, and generators, respectively. The OPF formulation is 
\begin{subequations}\label{opf}
\begin{align}
    \label{eq1}  &\min_{\substack{\vp^{g},\vq^{g}, \vp,\\ \vq,\boldsymbol{\theta},\boldsymbol{v}}} \quad  \sum_{i\in \mathcal{N}} f_i (p_i^g)\\ 
   \label{eq6} & \text{s.t.}  \quad \theta_i = 0 \text{ for } i \in \setS, \\ 
  \nonumber & \forall i \in \mathcal{N},\, \forall (i,j) \in \mathcal{E}: \\
   \label{eq2} & \quad p_i^g- p_i^d = \sum_{(i,j)\in \mathcal{E}} p_{ij} + g^{sh}_i v_i^2, \\
    \label{eq3} & \quad q_i^g- q_i^d = \sum_{(i,j)\in \mathcal{E}} q_{ij} - b^{sh}_i v_i^2, \\
   \label{eq4} \begin{split}
    & \quad p_{ij} = v_i^2G_{ij} \\
    & \quad \quad \quad \;\; -v_i v_j \big[G_{ij}\cos(\theta_{ij})  +B_{ij}\sin(\theta_{ij})\big] ,
   \end{split}\\
   \label{eq5} \begin{split}
      & \quad q_{ij} = -v_i^2(B_{ij}^{sh} + B_{ij})  \\ 
      & \quad \quad \quad -v_i v_j \big[G_{ij}\sin(\theta_{ij}) -B_{ij}\cos(\theta_{ij})\big], 
   \end{split}\\
   \label{eq7} & \quad \underline{P}_{i}^g \le p_i^g \le \overline{P}_{i}^g,    \\ 
   \label{eq8} & \quad \underline{Q}_{i}^g \le q_i^g \le \overline{Q}_{i}^g,  \\ 
   \label{eq9} & \quad \underline{V}_i \le v_i \le \overline{V}_i,   \\
  \label{eq10} & \quad p_{ij}^2 + q_{ij}^2 \leq \left(\overline{S}_{ij}\right)^2,   
\end{align}
\end{subequations}
\noindent where $f_i$ is the cost function and $p_i^g$, $q_i^g$ are the real and reactive power outputs, respectively, of generator $g\in\mathcal{G}$ located at bus $i \in \mathcal{N}$. We define $\theta_{ij} = \theta_i - \theta_j$ for $(i,j) \in \mathcal{E}$. The series conductance and susceptance of line $(i,j) \in \mathcal{E}$ are $G_{ij}$ and $B_{ij}$, while $\overline{S}_{ij}$ denotes the line's thermal limit. The shunt conductance and susceptance at bus~$i\in\mathcal{N}$ are $g^{sh}_i$ and $b^{sh}_i$. Each bus~$i\in\mathcal{N}$ has a voltage phasor $v_i\angle \theta_i$. We denote the real power demand at bus $i\in\mathcal{N}$ as $p_i^d$ and reactive power demand as $q_i^d$. Also, $\setS$ contains the reference bus.
The OPF problem minimizes the generation cost in~\eqref{eq1} subject to the AC power flow equations~\eqref{eq2}--\eqref{eq5}, the voltage limits and generators' power output limits~\eqref{eq7}--\eqref{eq9}, and the lines' thermal limits~\eqref{eq10}. Note also that we set the phase angle to 0 at a selected reference bus in \eqref{eq6}.

\subsection{Distributed Optimal Power Flow} \label{sec:dopf}
\begin{figure}[t]
    \centering 
    \includegraphics[height=5cm]{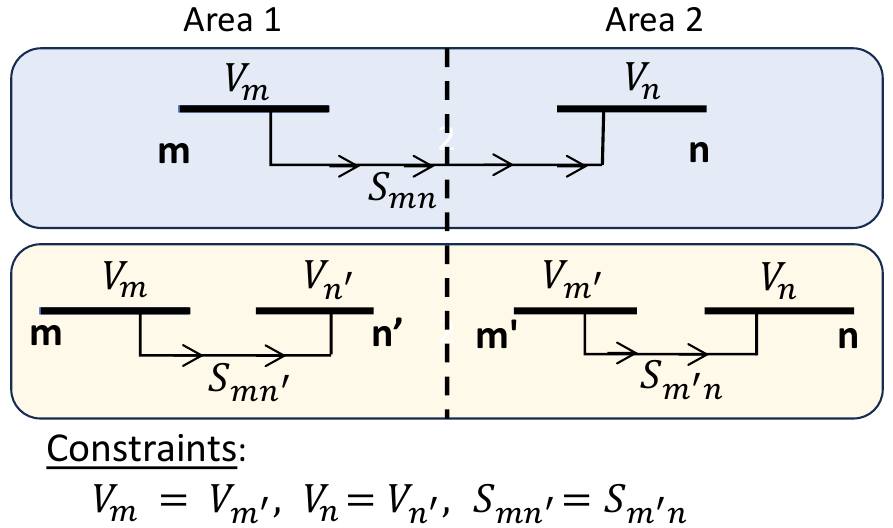} 
        \caption{Decomposition of power network}
        \label{fig:decomposition}
        \vspace{-1em}
\end{figure}

In the distributed OPF formulation, the power network is divided into regions, each under the control of a separate computing agent. When branch terminals are in different regions, we add fictitious buses as shown in Figure \ref{fig:decomposition} and set consistency constraints to ensure that the fictitious variables match the original variables in the neighbor's region. 
 
We can solve the distributed OPF formulation using alternating distributed algorithms (ADAs). In such algorithms, separate computing agents solve OPF subproblems over their region of the network. They augment their local OPF objective with relaxed consistency constraints using boundary variable values shared from neighboring agents. Agents iteratively solve their OPF subproblems and share boundary variable data until the consistency constraints are satisfied. This paper focuses on the ADMM algorithm, although our methods apply directly to other ADAs such as APP and ATC. 

We formulate the distributed OPF problem for ADMM as follows. Let $\mathcal{G}_m$, $\mathcal{N}_m$, and $\mathcal{E}_m$ denote the sets of generators, buses, and lines in region $m$, respectively. We denote the set of shared variables in region $m$ with $\mathcal{N}_m^s$. In addition, we use the same notation for variables as in \eqref{opf} but add dots to designate agents' copies of variables in their region, so that, e.g., $\dot{p}_{i,m}^g$ denotes region $m$'s copy of the power generation at bus $i$. The consistency constraints are relaxed with the augmented Lagrangian technique. The vector $\vz_m$ contains all shared variables in $\mathcal{N}_m^s$, and the vector $\bar{\vz}_m$ is a ``central'' variable which accounts for all neighbors' copies of the shared variables. In traditional ADMM, this variable is computed by a central coordinator, but for our formulation it simplifies to the average of the neighboring agents' shared variable values and is thus entirely separable, as in \cite{7428969}. At iteration $k$, agent~$m$ solves the following subproblem:
\begin{subequations}\label{dist_admm}
\begin{align}
\begin{split}
      &\min_{\substack{\dot{\vp}^{g,k},\dot{\vq}^{g,k}, \dot{\vp}^k,\\ \dot{\vq}^k,\dot{\boldsymbol{\theta}}^k,\dot{\boldsymbol{v}}^k,\vz^k_m}} \sum_{i\in \mathcal{N}_m} f_i (\dot{p}_{i,m}^{g,k}) + (\vy_m^{k-1})^T \vz_m^k \label{deq1} \\ 
   & \quad \quad \quad  \quad \quad + \frac{\alpha}{2} ||\vz_m^k - \bar{\vz}_m^{k-1}||_2^2  
\end{split}\\
  \label{deq10} & \text{s.t.} \quad \dot{\theta}_i = 0 \text{ for } i \in \setS, 
   \\
   \nonumber & \forall i\in \mathcal{N}_m, \forall (i,j)\in \mathcal{E}_m: \\
   & \quad \dot{p}_{i,m}^{g,k}- p_i^d = \sum_{(i,j)\in \mathcal{E}_m} \dot{p}_{ij,m}^k + g^{sh}_i (\dot{v}_{i,m}^k)^2, \label{deq2}\\
   & \quad \dot{q}_{i,m}^{g,k}- q_i^d = \sum_{(i,j)\in \mathcal{E}_m} \dot{q}_{ij,m}^k - b^{sh}_i (\dot{v}_{i,m}^k)^2, \label{deq3}\\
   \begin{split}
    & \quad \dot{p}_{ij,m}^k = (\dot{v}_{i,m}^k)^2G_{ij}  \\
    & \quad \quad \quad \;\; -\dot{v}_{i,m}^k v_j^k \big[G_{ij}\cos(\theta_{ij}^k)+B_{ij}\sin(\dot{\theta}_{ij,m}^k)\big],  \label{deq4}
   \end{split}\\
   \begin{split}
    & \quad \dot{q}_{ij,m}^k = -(\dot{v}_{i,m}^k)^2(B_{ij}^{sh} + B_{ij}) \\   
   & \quad \quad \quad-\dot{v}_{i,m}^k v_j^k \big[G_{ij}\sin(\dot{\theta}_{ij,m}^k) -B_{ij}\cos(\dot{\theta}_{ij,m}^k)\big], \label{deq5}
   \end{split}\\
      & \quad \underline{P}_{i}^g \le \dot{p}_{i,m}^{g,k} \le \overline{P}_{i}^g, \label{deq6} \\ 
    & \quad \underline{Q}_{i}^g \le \dot{q}_{i,m}^{g,k} \le \overline{Q}_{i}^g, \label{deq7} \\
   & \quad \underline{V}_i \leq \dot{v}_{i,m}^k \le \overline{V}_i,  \label{deq8} \\
   & \quad (\dot{p}_{ij,m}^k)^2 + (\dot{q}_{ij,m}^k)^2 \leq \left(\overline{S}_{ij}\right)^2. \label{deq9} 
\end{align}
\end{subequations}
Note that $\alpha$ is a user-defined penalty parameter. After solving \eqref{dist_admm}, each agent $m$ shares the boundary variable values $\vz_m$ with their neighbors. Then, each agent $m$ updates the $\bar{\vz}_m$ variables. For every neighbor $n$ of agent $m$, there is a set of variables $\mathcal{N}_{m,n}^s$ that are shared between agents $m$ and $n$. We denote agent $m$'s copies of these shared variables as the vector $\vz_{m,n}$ and agent $n$'s copies of these shared variables as the vector $\vz_{n,m}$. Agent $m$ updates the average of local shared variables and shared variables received from neighbor $n$, $\bar{\vz}_{m,n}$, as
 \begin{equation}
     \label{update} 
     \bar{\vz}_{m,n}^k = \frac{1}{2}(\vz_{m,n}^k + \vz_{n,m}^k).
 \end{equation}
Each agent $m$ updates their Lagrange multipliers as 
 \begin{equation}\label{lagrange}
     \vy_{m}^{k}=\vy_{m}^{k-1} + \alpha(\vz_m^{k}-\bar{\vz}_m^{k}).
 \end{equation}
 Thus, the iterative algorithm alternates between minimizing the agents' subproblems in \eqref{dist_admm}, updating the average copies of variables shared between agents in \eqref{update}, and updating dual variables in \eqref{lagrange}. Typically, the stopping criterion is based on primal and dual residuals \cite{admm_book}. The vector of primal residuals $r^k$ contains the difference between local and central copies of all boundary variable values:
 \begin{equation}\label{pr_res}
     r^k = \begin{bmatrix} \vz_1^T - \bar{\vz}_1^T & \vz_2^T - \bar{\vz}_2^T & ... & \vz_M^T - \bar{\vz}_M^T \end{bmatrix}^T. 
 \end{equation}
 The dual residual is 
 \begin{equation}\label{du_res}
     s^k =  -\alpha (\bar{\vz}^{k} - \bar{\vz}^{k-1}),
 \end{equation}
where we have collected all central copies of boundary variables into one vector $\bar{\vz}$. 
The algorithm terminates when the primal and dual residual norms fall below the respective primal and dual tolerances:
\begin{equation}\label{termination}
    ||r^k|| \leq \epsilon^{pri}, \quad ||s^k|| \leq \epsilon^{dual}.
\end{equation}
The next section discusses how these tolerances are selected. 

\section{Selecting Convergence Tolerances}\label{sec:select_tol}
We terminate the distributed OPF algorithm when the primal and dual residuals are sufficiently small. The most widely referenced work on ADMM, \cite{admm_book}, suggests using the $\ell_2$-norm of the primal and dual residuals as the stopping criterion. Many papers on distributed AC OPF also use the $\ell_2$-norm of both primal and dual residuals \cite{6748974, 7440858, zeng_kody_kim_kim_molzahn-rl_for_dist_opt, 8573893}. Other papers use the $\ell_\infty$- or $\ell_2$-norm of the dual residuals only \cite{7244261,8409328}, while yet other publications use the $\ell_\infty$- or $\ell_2$-norm of the primal residuals \cite{7776940,780896,alkhraijah_menendez_molzahn-distributed_optimization_nonideal}. Most of the above works select a tolerance in the range of $[10^{-5},10^{-3}]$, although \cite{admm_book} proposes a method to define tolerances based on the scale of the variables:
\begin{align}\label{epsilons}
\begin{split}
    & \epsilon^{pri} = \sqrt{p}\epsilon^{abs} + \epsilon^{rel}\max \{||\mA \vx^k||_2, ||\mB \vz^k||_2, ||\vc||_2\}, \\ 
    & \epsilon^{dual} = \sqrt{n}\epsilon^{abs} + \epsilon^{rel} ||\mA^T \vy^k||_2, 
\end{split}
\end{align}
where $\epsilon^{abs}$, $\epsilon^{rel}$ are user-selected absolute and relative tolerances, respectively, and the notation is for a general ADMM formulation which minimizes a function $f(\vx) + g(\vz)$ subject to the coupling constraint $\mA \vx + \mB \vz = \vc$. In~\eqref{epsilons}, $p$ is the number of constraints and $n$ is the number of shared variables. 

Our analysis will focus on $\epsilon^{pri}$, and we will determine convergence based on the primal residuals alone. The requirement for small primal residuals, $||r^k|| \leq \epsilon^{pri}$, results in near feasibility of the final solution by satisfying consistency constraints. The requirement for small dual residuals, $||s^k|| \leq \epsilon^{dual}$, is related to optimality of the final solution. This paper's analysis is primarily concerned with feasibility, and our methods are designed to ensure feasible solutions for a given choice of $\epsilon^{pri}$. However, our numerical results demonstrate that in practice, given appropriate choice of penalty parameter $\alpha$, setting the stopping criterion based on primal residuals results in solutions that are both nearly optimal and nearly feasible. 

We choose the $\ell_\infty$ norm as the convergence criterion in our analyses for two reasons. First, the $\ell_\infty$ norm of the shared variable mismatches is immediately interpretable as the maximum variable mismatch and has units of p.u. for voltage magnitudes and power flows and radians for voltage angles. Second, it allows for simple linear constraints in the worst-case violation optimization problem we formulate in Section~\ref{sec:analysis_and_algorithms}. Extensions of the algorithm we will propose in this paper to other norms are conceptually straightforward.

\begin{figure}
     \centering
     \begin{subfigure}[b]{\linewidth 
     }
         \centering
         \includegraphics[height=4.5cm]{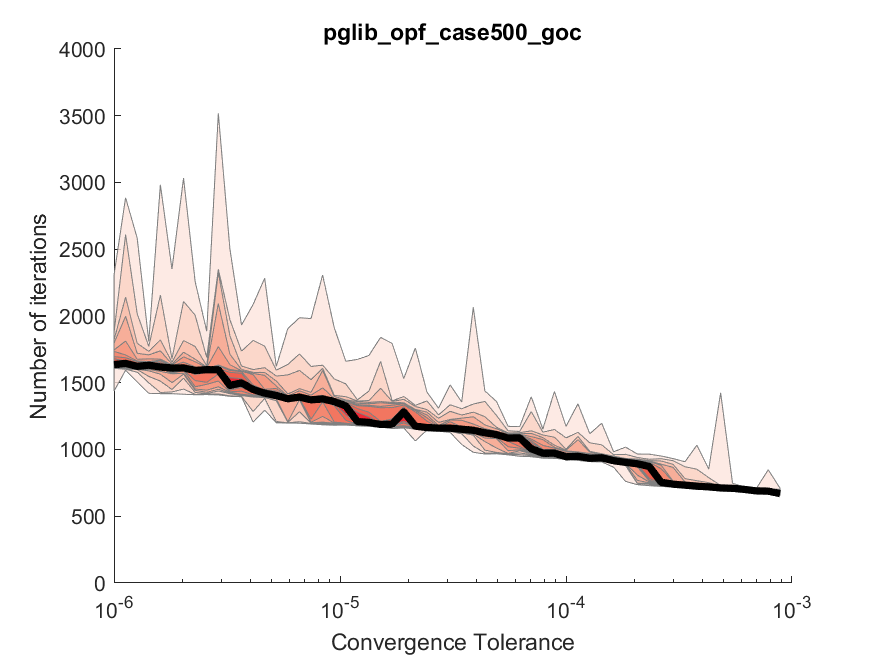}
         \caption{Num. iterations to converge vs. convergence tolerance}
         \label{fig:iter_vs_conv}
     \end{subfigure}
     \hfill
     \begin{subfigure}[b]{\linewidth}
         \centering
         \includegraphics[height=4.5cm]{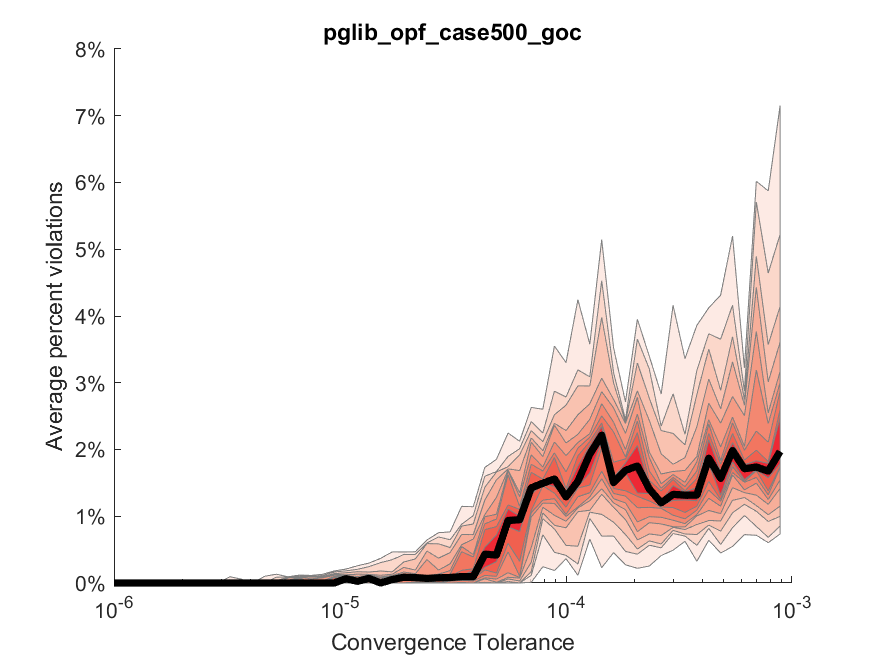}
         \caption{Average percent violation vs. convergence tolerance}
         \label{fig:viol_vs_conv}
     \end{subfigure}
     \hfill
     \begin{subfigure}[b]{\linewidth}
         \centering
         \includegraphics[height=4.5cm]{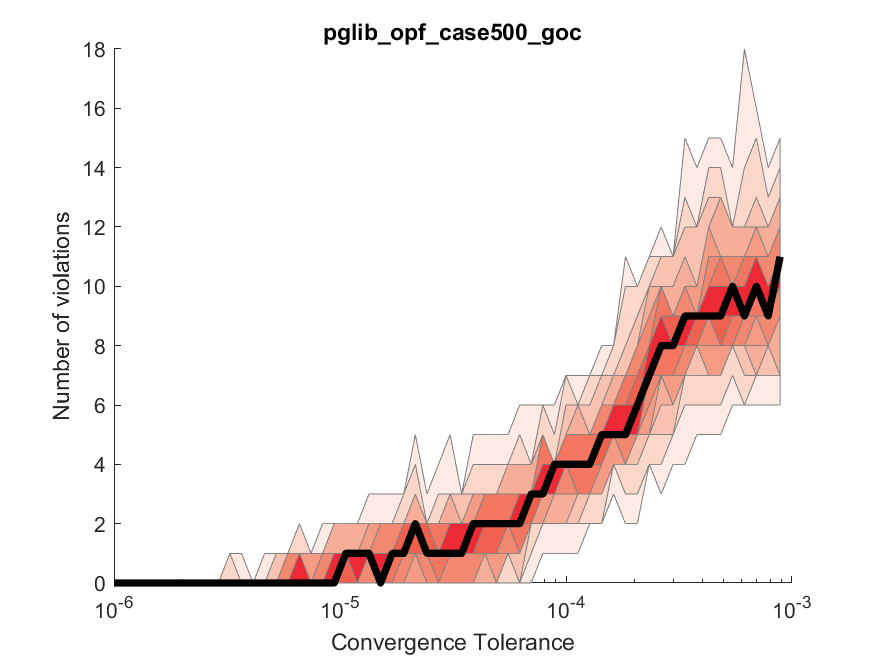}
         \caption{Num. constraint violations vs. convergence tolerance}
         \label{fig:numviol_vs_conv}
     \end{subfigure}
        \caption{Impact of convergence tolerance}
        \label{fig:impact_conv_tol}
\end{figure}

Changing the convergence tolerance $\epsilon^{pri}$ impacts the speed, feasibility, and optimality of distributed optimization algorithms. We provide an illustrative example using the 500-bus test case from the PGLib-OPF archive~\cite{ieee_pes_pglib-opf_task_force_power_2019} divided into 8 regions for distributed optimization. We use the \mbox{PowerModelsADA} library~\cite{alkhraijah_harris_coffrin_molzahn-pmada} to solve the distributed OPF problem using the ADMM algorithm. We run the distributed OPF algorithm 2000 times, sweeping the convergence tolerance $\epsilon^{pri}$ from $10^{-6}$ to $10^{-3}$, and each time randomly perturbing loads by selecting values between $70\%$--$130\%$ of nominal. Once the distributed OPF algorithm terminates, we run an AC power flow on the system using the distributed OPF generator dispatch and determine if the results violate any bounds on voltage magnitudes, reactive power generation, or line flows. The results are shown in Figure \ref{fig:impact_conv_tol}, where the shaded red bands around the median line in black show every fifth percentile of the results. Figure \ref{fig:iter_vs_conv} shows that the number of iterations required to reach convergence decrease significantly as $\epsilon^{pri}$ increases. Figure \ref{fig:viol_vs_conv} shows the average percent violation for the constraint violations that occur, where we define the average percent violations as 
\begin{equation}\nonumber
\frac{1}{N_v} \sum_{i \in \setC} \frac{\max \{ x_i^{AC-PF} - x_i^{max}, x_i^{min} - x_i^{AC-PF}, 0 \}}{x_i^{max}-x_i^{min}},
\end{equation}
where $N_v$ is the number of violated constraints and $\setC$ contains indices of all variables representing voltage magnitudes, reactive power injections, and line flows. We denote the value of the $i$-th variable computed by the AC power flow as $x_i^{AC-PF}$, and its minimum and maximum values as $x_i^{min}$ and $x_i^{max}$. The median number of violations per run is shown in Figure~\ref{fig:numviol_vs_conv}. As the maximum shared variable mismatches approach $10^{-4}$, the power flow solution from the distributed OPF operating point starts to have non-negligible constraint violations, which increase with larger tolerances $\epsilon^{pri}$. This behavior is exactly what we would expect, since as $\epsilon^{pri}$ becomes sufficiently large, the consistency constraints for boundary variables are not satisfied and the distributed OPF solution may not be feasible. Note that while the computation time decreases at an approximately linear rate, there is a sudden steep increase in the average percent violations at about $\epsilon^{pri} = 4\times 10^{-5}$. 

\section{Analysis and Bound Tightening Algorithm}\label{sec:analysis_and_algorithms}
As shown by the example in the prior section, sufficiently loose convergence tolerances may lead to non-negligible constraint violations. This motivates the development of techniques for bounding the worst-case constraint violations and mitigating their impacts on the resulting solutions. We develop a method to determine the worst-case constraint violations that may occur from applying the distributed OPF solution converged to a given tolerance $\epsilon^{pri}$ to the system. We first formulate an optimization problem which finds the worst-case constraint violations for a given maximum boundary variable mismatch $\epsilon^{pri}$. Next, we propose a bound tightening algorithm which alternates between finding the worst-case violations and subsequently tightening the constraints to mitigate those violations. Provided that the true worst-case violation is found for each constraint, the distributed OPF algorithm run on the bound-tightened case will not violate any original constraints once applied to the system. 

\begin{figure}[t]
    \centering 
    \includegraphics[width=0.95\linewidth]
    {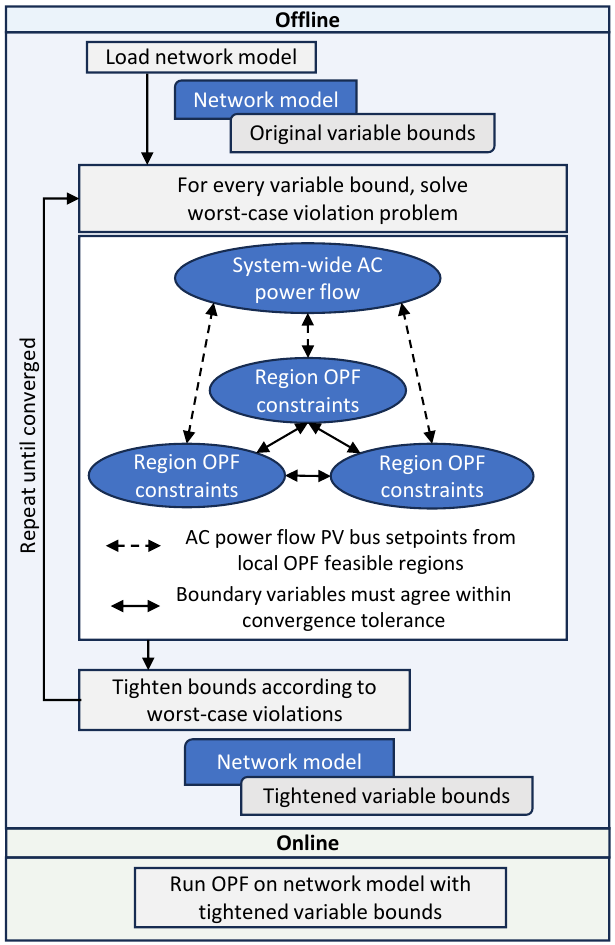} 
        \caption{Bound tightening algorithm overview}
        \label{fig:flow_chart}
        \vspace{-2em}
\end{figure}

The worst-case violation analysis and constraint-tightening algorithm is useful for OPF problems solved repeatedly, with a constant network model and loads varying with each run. The proposed algorithm requires the ability to perform offline calculations where information regarding the entire system is available. Offline, we formulate an optimization problem which finds the worst-case violation, allowing the loads to take any values within a specified range, given some $\epsilon^{pri}$. We iterate between solving the worst-case violation problem for all variable bounds and tightening the bounds according to the worst-case violations until the algorithm converges. We show an overview of the full bound tightening algorithm in Figure \ref{fig:flow_chart} and next provide the formulation and algorithm details.

\subsection{Notation and Modeling Choices}\label{sec:notation_modeling}
We propose a method to determine the worst-case constraint violations that may occur for a convergence tolerance $\epsilon^{pri}$. We consider a network model with buses in $\mathcal{N}$, generators in $\mathcal{G}$ and lines in  $\mathcal{E}$. We use the same notation for system variables as in Section~\ref{sec:opf}. Note that while the active and reactive power demands $p_i^d$ and $q_i^d$ at each bus~$i$ are fixed for the OPF formulation~\eqref{opf} in Section~\ref{sec:opf}, this worst-case violation problem has the power demands as variables. We allow the power demands to vary by a factor~$r$; for example, if $r = 0.5$, then the active and reactive power demands may take any value between 50\%--150\% of their nominal values, denoted as $p_i^{d,nom}$ and $q_i^{d,nom}$. We keep a constant power factor by modeling consistent perturbations to both active and reactive power at a given bus by the same factor $r$. With this approach, we can perform offline computations for the tightened variable bounds without needing information regarding the exact values of loads that would only be available to local agents in real-time calculations, as shown in Figure \ref{fig:flow_chart}. We denote $\setN_g$ as the set of buses with generators and $\setS$ as the slack bus. 

We use the same notation for variables contained in local regions as in Section~\ref{sec:dopf}. We put a dot over variables belonging to local systems to distinguish them from the central system variables. Again, $\setM$ denotes the set of agents controlling regions in the distributed OPF problem and $\setA_m$ denotes the neighbors of agent $m$. Also, the vector $\vz_{m,n}$ contains agent $m$'s copies of all boundary variables shared between agents $m$ and $n$, which includes voltage magnitudes and angles for boundary buses and active and reactive power flows on boundary lines. We denote the amount by which original bounds have been tightened by $\lambda_{\underline{V}_i},\lambda_{\overline{V}_i}$ for lower and upper bounds on the voltage at bus $i$, $\lambda_{\underline{Q}_i},\lambda_{\overline{Q}_i}$ for lower and upper bounds on reactive power generation at bus $i$, and $\lambda_{\overline{S}_{ij}}$ for the upper bound on apparent power flow across line $(i,j)$. For instance, with a constraint tightening of $\lambda_{\overline{V}_i}$, the upper voltage limit~\eqref{deq8} in an agent's subproblem becomes $\dot{v}_{i,m}^k \leq \overline{V}_i  - \lambda_{\overline{V}_i}$. We collect the amount of bound tightening on all variables into one vector~$\vlambda$. 

We note that the worst-case violation on any variable bound depends on the choice of convergence tolerance $\epsilon^{pri}$ and on the amount of bound tightening $\vlambda$. Therefore, we denote the worst-case violations on upper and lower bounds on voltage magnitudes at bus $i$ as $W_{\overline{v}_i}(\epsilon^{pri},\vlambda)$ and $W_{\underline{v}_i}(\epsilon^{pri},\vlambda)$, respectively; on upper and lower bounds on reactive power generation at bus $i$ as $W_{\overline{q}_i}(\epsilon^{pri},\vlambda)$ and $W_{\underline{q}_i}(\epsilon^{pri},\vlambda)$, respectively; and on upper bounds on line apparent power flows at line $(i,j)$ as $W_{\overline{s}_{ij}}(\epsilon^{pri},\vlambda)$. We next describe an optimization formulation for calculating the worst-case constraint violations for a given convergence tolerance $\epsilon^{pri}$ and bound tightening $\vlambda$.

\subsection{Worst-Case Violation Formulation}\label{sec:viol_formulation}
We next formulate an optimization problem that computes the worst-case constraint violations for a given range of load variation and convergence tolerance. To formulate this problem, we begin with constraints that belong to two categories:
\begin{enumerate}
    \item \emph{Distributed OPF constraints} which represent the behavior of the distributed OPF algorithm. The variables kept by local regions (recall that these are marked with a dot) must satisfy OPF constraints within that region. In addition, consistency constraints require that the differences between neighboring regions' copies of shared variables are no more than $\epsilon^{pri}$.
    \item \emph{System-wide AC power flow constraints} which represent the physical behavior of the system under a distributed OPF solution dispatch. These constraints involve variables representing the physical system (which are not marked with a dot) and are the traditional AC power flow equations. The setpoints for PV buses in the AC power flow come from the distributed OPF variable values. 
\end{enumerate}

To compute worst-case violations, we will form optimization problems that have the following constraints:
\begin{subequations}\label{constraints}
\begin{align} 
    \nonumber & \forall m \in \setM : \\ 
   \label{ceq1} & \quad \quad  \eqref{deq2}\text{--}\eqref{deq6}  \\
   \label{ceq2} & \quad \quad  \underline{V}_i + \lambda_{\underline{V}_i} \leq \dot{v}_{i,m} \le \overline{V}_i - \lambda_{\overline{V}_i}, \quad \forall i \in \setN_m, \\ 
   \label{ceq3} & \quad \quad  \underline{Q}_{i}^g + \lambda_{\underline{Q}_i} \le \dot{q}_{i,m}^{g} \le \overline{Q}_{i}^g - \lambda_{\overline{Q}_i}, \quad \forall i \in \setN_m,\\
   \label{ceq4} & \quad \quad (\dot{p}_{ij,m})^2 + (\dot{q}_{ij,m}^k)^2 \leq \left(\overline{S}_{ij} - \lambda_{\overline{S}_{ij}}\right)^2,\; \forall (i,j) \in \setE_m,  \\
  \label{ceq5} & \quad \quad  ||\vz_{m,n} - \vz_{n,m}||_\infty \leq \epsilon^{pri}, \quad \forall n \in \setA_m, \\
   \label{ceq6} & p_i^{d} = p_i^{d,nom} + \tilde{p}_i, \quad |\tilde{p}_i| \leq r\cdot p_i^{d,nom} ,  \quad  \forall i \in \setN, \\ 
   \label{ceq7} & q_i^{d} = q_i^{d,nom} + \tilde{q}_i, \quad |\tilde{q}_i| \leq r\cdot q_i^{d,nom}, \quad  \forall i \in \setN,  \\ 
    \label{ceq8} & p_i^g = \dot{p}_{m,i}^g, \quad v_i = \dot{v}_{m,i},  \quad \forall i \in \setN_g,  \\ 
   \label{ceq9} & v_i = \dot{v}_{m,i}, \quad \theta_i = \dot{\theta}_{m,i}, \quad \text{ for } i \in \setS, 
    \\ 
   \label{ceq10} & \forall i \in \mathcal{N},\, \forall (i,j) \in \mathcal{E}: \nonumber \\ 
   & \quad \quad  p_i^g - p_i^d = \sum_{(i,j)\in \mathcal{E}} p_{ij} + g^{sh}_i v_i^2, \\
   \label{ceq11}& \quad \quad  q_i^g- q_i^d = \sum_{(i,j)\in \mathcal{E}} q_{ij} - b^{sh}_i v_i^2, \\
  \label{ceq12} \begin{split}
    & \quad \quad p_{ij} = v_i^2G_{ij} \\
    & \quad \quad \quad  \quad \;\; -v_i v_j \big[G_{ij}\cos(\theta_{ij})  +B_{ij}\sin(\theta_{ij})\big], 
   \end{split}\\
   \label{ceq13} \begin{split}
      & \quad \quad  q_{ij} = -v_i^2(B_{ij}^{sh} + B_{ij})  \\ 
      & \quad \quad \quad \quad -v_i v_j \big[G_{ij}\sin(\theta_{ij}) -B_{ij}\cos(\theta_{ij})\big],
   \end{split} \\ 
   \label{ceq14} & \quad \quad  v_i \geq \underline{V}. 
\end{align}
\end{subequations}
Constraint \eqref{ceq1} ensures that the solution from each agent's region satisfies the power balance and line power flow constraints in that region. Constraints \eqref{ceq2}--\eqref{ceq4} are the voltage magnitude, reactive power injection, and line apparent power flow bounds imposed on variables in each region's OPF problem. Constraint \eqref{ceq5} ensures that the maximum boundary variable mismatch is not greater than $\epsilon^{pri}$ to model the agents reaching their convergence tolerances. Constraints \eqref{ceq6}--\eqref{ceq7} set the amount by which loads may vary as described in Section~\ref{sec:notation_modeling}.\footnote{Note that \eqref{ceq1}--\eqref{ceq5} enforce additional implicit constraints on the loads since some loading conditions within the variability allowed by $r$ may not be feasible given each region's OPF constraints and the requirement that neighboring regions' shared variables agree to within a tolerance of $\epsilon^{pri}$. If a loading condition is not feasible for \eqref{constraints}, then it is not feasible for the original OPF problem \eqref{opf}, so it is acceptable for \eqref{constraints} to exclude these infeasible loading points.} Constraint \eqref{ceq8} sets the system active power injection and voltage magnitude variables for PV buses to the setpoints from the distributed OPF solution. Constraint \eqref{ceq9} sets the slack bus voltage angle to 0 and the voltage magnitude to the result from the distributed OPF solution. Constraints \eqref{ceq10}--\eqref{ceq13} are the traditional AC power balance and line flow constraints for the system. These represent physical system behavior under the distributed OPF dispatch, where the active power injection and voltage magnitudes at PV buses are set to the results of the distributed OPF computation. Constraint \eqref{ceq14} is designed to prevent the solver from finding a low-voltage solution to the AC power flow equations in \eqref{ceq9}--\eqref{ceq13} by providing a lower bound for the voltage magnitudes.\footnote{The value of $\underline{V}$ is chosen to be much lower than the lowest anticipated voltage (e.g., 0.7 per unit) so that the only effect of~\eqref{ceq14} is avoiding a low-voltage power flow solution for~\eqref{ceq10}--\eqref{ceq13}.} 

We add an appropriate objective to \eqref{constraints} to find the worst-case violations of bounds on voltage magnitudes, reactive power injections, and line apparent power flows. For example, to compute the worst-case violation of the upper voltage limit at bus $i$ for a given convergence tolerance of $\epsilon^{pri}$ and bound tightening values $\vlambda$, we first solve
\[\overline{v}_i^* = \max\, v_i \text{ subject to } \eqref{constraints}. \]
The worst-case violation is then
\[W_{\overline{v}_i}(\epsilon^{pri},\vlambda) = \overline{v}_i^* - \overline{V}_i. \]
Similarly, for the lower bound on voltage magnitude at bus~$i$, we first solve 
\[\underline{v}_i^* = \min\, v_i \text{ subject to } \eqref{constraints}. \]
The worst-case violation of the lower voltage bound is then
\[W_{\underline{v}_i}(\epsilon^{pri},\vlambda) = \underline{V}_i - \underline{v}_i^*. \]
Note that if we find $W_{\overline{v}_i}(\epsilon^{pri},\vlambda),W_{\underline{v}_i}(\epsilon^{pri},\vlambda) \leq 0$, then even in the worst case there is no violation of the bound constraint. 

Similarly, we maximize and minimize the variable $q_i^g$ at bus $i$ to compute worst-case violations of reactive power generation limits. For worst-case violations of apparent power flow limits on line $(i,j)$, we maximize $p_{ij}^2 + q_{ij}^2$ and then compute $W_{\overline{s}_{ij}}(\epsilon^{pri},\vlambda) = \sqrt{(p_{ij}^*)^2 + (q_{ij}^*)^2} - \overline{S}_{ij}$.

\subsection{Discussion}\label{sec:discussion}
Our formulation neglects the fact that the generator dispatch $\dot{p}_{m,i}^g$, $\dot{v}_{m,i}$ from the distributed OPF would optimally solve \eqref{dist_admm} for each agent $m$. Instead of requiring that the distributed OPF variables are \emph{optimal} for their region's OPF problem with relaxed consistency constraints, we require only that the distributed OPF variables are \emph{feasible} for the region's OPF problem. Thus, our formulation may be conservative, i.e., return worst-case violations larger than would actually be produced by the distributed optimization algorithm. An alternative formulation that enforces optimality of each region's OPF problems would lead to a computationally challenging bilevel problem. Our future work includes leveraging the optimality of $\dot{p}_{m,i}^g$, $\dot{v}_{m,i}$ to find less conservative worst-case violations. 

We also note that the non-convex nature of the worst-case violation constraints means that a solver may find a local, rather than global, solution and thus not identify the actual largest possible violation. Alternatively, one could form a variant of~\eqref{constraints} with relaxed AC power flow constraints~\cite{molzahn_hiskens-fnt2019}. The violation obtained by optimizing over a convex relaxation of the AC power flow equations will be equal to or greater than the actual largest possible violation. We chose to use the nonlinear AC power flow equations despite the possibility of local optima because problems constrained by convex relaxations may be slower to solve and may require careful implementation to ensure the relaxation is tight enough to avoid overly conservative bounds. We demonstrate via our results in Section~\ref{sec:results} that although a nonlinear programming solver may occasionally return a local solution, we observe no violations in practice when running distributed OPF on test cases with bounds tightened using the formulation with AC power flow equations. This suggests that local solvers perform well for our purposes.

We also assume that there is at most one relevant solution to the AC power flow equations \eqref{ceq10}--\eqref{ceq13} for all power injections within the specified range. Although there may be many ``low-voltage'' solutions, typically there is only one ``high-voltage'' solution with near-nominal voltage magnitudes, and this high-voltage solution is the one we desire to find. We add constraint \eqref{ceq14} to screen out low-voltage power flow solutions. We note that the modeling challenges associated with nonconvexities and low-voltage power flow solutions are similar to those faced in stochastic and robust optimization problems; see~\cite[Section~XI]{roald_pozo_papavasiliou_molzahn_kazempour_conejo-pscc2022_survey} for further discussion. 

\subsection{Bound Tightening}

As demonstrated for a representative test case in Figure~\ref{fig:iter_vs_conv}, choosing a larger convergence tolerance $\epsilon^{pri}$ can dramatically decrease the number of iterations for the distributed optimization algorithm. However, larger tolerances may also result in constraint violations due to the inconsistency between neighboring regions' copies of boundary variables. We propose a method to tighten constraints such that the dispatch from the distributed OPF algorithm when converged to a given $\epsilon^{pri}$ is guaranteed not to violate the original constraints. Although the setting and application is different, our alternating algorithm is conceptually similar to those proposed in \cite{8017474,8442889}, which use constraint tightening to make AC OPF problems robust to uncertainty in power demand or generation. 

\begin{figure}[t]
\centering
\resizebox{\linewidth}{!}{%
\begin{tikzpicture}[scale=0.880,node distance=8mm]
    

    \node[shape=rectangle, draw, align=center,top color=white] (initial) at (3,0) 
    {Initialize: $k=0$ and bound tightenings $\vlambda^0 = \vzero$.};
    
    \node[shape=rectangle, draw, align=center,top color=white] (opf) at (3,-2) {Compute worst-case violations for all bounds: \\ $W_{\overline{v}_i}(\epsilon^{pri},\vlambda^{k})$, $W_{\underline{v}_i}(\epsilon^{pri},\vlambda^{k})$ for all $i \in \setN $, \\
    $W_{\overline{q}_i}(\epsilon^{pri},\vlambda^{k})$, $W_{\underline{q}_i}(\epsilon^{pri},\vlambda^{k})$ for all $i \in \setN_g $, \\
    $W_{\overline{s}_{ij}}(\epsilon^{pri},\vlambda^{k})$ for all $(i,j) \in \setE $.};
    
    
    \node[shape=rectangle, draw, align=center,top color=white] (tightening) at (3,-4.3) {Compute constraint tightenings $\vlambda^{k+1}$ \\ according to Algorithm 1.};

    \node[shape=rectangle, draw, align=center,top color=white] (feasibility) at (3,-5.9) {Solve AC OPF with nominal loads. \\ Is there a feasible solution? };

    \node[shape=rectangle, draw, align=center,top color=white] (convergence) at (3,-7.7) {Check convergence: \\ Is $||\vlambda^{k+1} - \vlambda^k||_2 \leq \Gamma$ ?};

    \node[shape=rectangle, draw, align=center,top color=white] (yes) at (7,-9.6) {Save final tightenings. \\ Bound tightening completed for $\epsilon^{pri}$.};

    \node[shape=rectangle, draw, align=center,top color=white] (not_feasible) at (0,-9.6) {Bound tightening failed for $\epsilon^{pri}$.};

    \node[shape=rectangle, draw, align=center,top color=white] (no) at (8,-5.7) {$k \leftarrow k+1$};

    \draw[->,line width=1.5pt] (initial) -- (opf);
    \draw[->,line width=1.5pt] (opf) -- (tightening);
    \draw[->,line width=1.5pt] (tightening) -- (feasibility);
    \draw[->,line width=1.5pt] (feasibility) -- (convergence);
    \draw[->,line width=1.5pt] (convergence) -- (yes);
    \draw[->,line width=1.5pt] (convergence) edge[bend right=20] (no);
    \draw[->,line width=1.5pt] (no) edge[bend right=20] (opf);
    \draw[->, line width=1.5pt] ([xshift=-2cm]feasibility.south) -- ([xshift=-1.5cm]not_feasible.north);
    \node[draw=none,fill=none] (yeslabel) at (4.3,-8.7) {Yes};
    \node[draw=none,fill=none] (yeslabel1) at (2.6,-6.75) {Yes};
    \node[draw=none,fill=none] (nolabel) at (5.7,-7.8) {No};
    \node[draw=none,fill=none] (nolabel1) at (0.1,-6.9) {No};
	    
\end{tikzpicture}
}
\caption{Alternating algorithm for robust AC~OPF problems}
\label{alg:iteration}
\vspace*{-1em}
\end{figure}
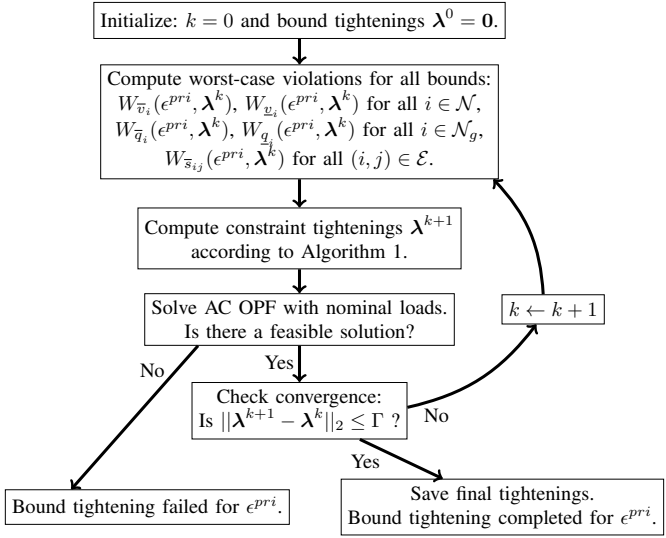

We now present the bound tightening algorithm. We show the steps of the algorithm in Figure~\ref{alg:iteration}. 
First, we initialize the tightening for each constraint to 0 by setting $\vlambda^0 = \vzero$. Second, we compute worst-case violations of all bounds. Third, we compute updated tightening values $\vlambda^k$ based on these violations as shown in Algorithm \ref{alg:tightening}. Note that we use $s$ as a generic variable index and observe that the update of $\lambda_s$ follows the same logic for tightening of upper bounds $\lambda_{\overline{V}_i}$, $\lambda_{\overline{Q}_i}$, $\lambda_{\overline{S}_{ij}}$ and tightening of lower bounds $\lambda_{\underline{V}_i}$, $\lambda_{\underline{Q}_i}$.
\begin{algorithm}[b]
\caption{Updating constraint tightenings $\lambda_s^k$}\label{alg:tightening}
\begin{algorithmic}
    \If {$W_{s}(\epsilon^{pri},\vlambda^{k-1}) > 0$}
    \State $\lambda_s^k = \lambda_s^{k-1} + W_{s}(\epsilon^{pri},\vlambda^{k-1})$
    \ElsIf {$\lambda_s^{k-1} > 0$}
    \State $\lambda_s^k = \lambda_s^{k-1} - \min \{  -W_{s}(\epsilon^{pri},\vlambda^{k-1}), \lambda_s^{k-1} \}$
    \EndIf 
\end{algorithmic}
\end{algorithm}
For a positive worst-case violation $W_r$, we increase the amount of tightening by $W_r$. We also check for unnecessary tightening: if the worst-case violation $W_r$ is negative (that is, the variable is within its bound) and there has already been some tightening so that $\lambda_r > 0$, we reduce the amount of tightening by $W_r$ or until $\lambda_r = 0$. Fourth, we solve an AC OPF problem on the system with nominal loads and bounds tightened by $\vlambda^{k+1}$ to make sure that the updated tightenings do not make the problem infeasible. Last, we evaluate the change in $\vlambda$ since the last iteration (as measured by the 2-norm) and return to Step 2 if this change is above a specified threshold $\Gamma$. Otherwise, the algorithm ends.


\subsection{Budget Uncertainty Set}
\begin{figure}[t]
    \centering 
    \includegraphics[height=5cm]{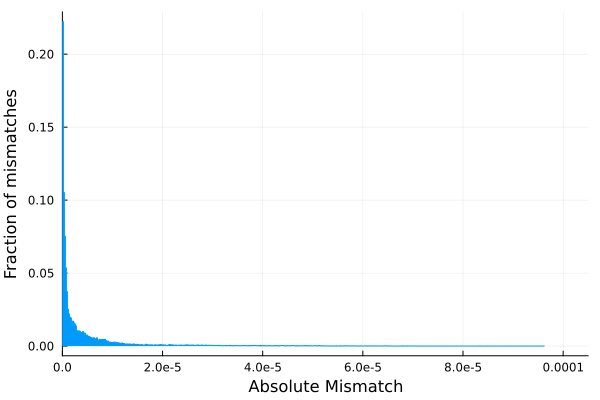} 
        \caption{Distribution of mismatches for case500 divided into eight regions, with distributed OPF converged to $\epsilon^{pri} = 10^{-4}$}
        \label{fig:mm_hist_118}
        \vspace{-1em}
\end{figure}

In the formulation \eqref{constraints}, we allow every shared variable to take on its maximum possible mismatch $\epsilon^{pri}$. However, in practice, agents reach consensus on some boundary variables more quickly than others. By the time the algorithm converges with a  maximum mismatch below $\epsilon^{pri}$, many of the other mismatches are much smaller than $\epsilon^{pri}$. We show a representative case in Figure~\ref{fig:mm_hist_118} and observe that most of the mismatches are much smaller than the convergence tolerance of $\epsilon^{pri} = 10^{-4}$. 

This motivates introducing the concept of budget uncertainty, which allows us to bound the total mismatch across the system and thus make less conservative predictions of the worst-case constraint violations. The budget uncertainty concept we use is similar to that used in~\cite{8016417}, although our ``uncertainty'' regards the mismatch in shared variable values in a mathematical distributed optimization problem, rather than coming from renewable power fluctuations. To add the uncertainty budget to our problem, we choose the budget size $\beta$ and augment \eqref{constraints} with the following constraint:
%
\begin{align} \label{budget_constraints}
   & \sum_{(m,n) \in \setP} \sum_{i \in \setI_{m,n}} |z^i_{m,n} - z^i_{n,m}| \leq \beta N_b \epsilon^{pri} \hspace{-1.85em}
\end{align}
where $\setP$ is the set of all neighboring agent pairs $(m,n)$, and the set $\setI_{m,n}$ contains indices for the specific variables shared between agents $m$ and $n$. Here, $z^i_{m,n}$ is agent $m$'s copy of the $i$-th boundary variable shared between agents $m$ and $n$. The total number of boundary variables in the system is 
\begin{equation} \nonumber  
N_b = \sum_{(m,n) \in \setP} |\setI_{m,n}|.
\end{equation}

Without adding \eqref{budget_constraints}, the constraints \eqref{constraints} allow for the total mismatch in the system, i.e., the sum of all boundary variable mismatches, to reach $N_b \epsilon^{pri}$, because every boundary variable can reach a mismatch of $\epsilon^{pri}$. We add the bounds on the total mismatch in \eqref{budget_constraints} so that the sum of absolute mismatches across the system is no more than a fraction $\beta$ of $N_b \epsilon^{pri}$. Note that it is straightforward to reformulate \eqref{budget_constraints} as a set of linear inequalities, which is how we implemented this constraint. 

The choice of parameter $\beta$ allows us to control the conservativeness of the constraint tightenings $\vlambda$. With $\beta < 1$, we cannot guarantee finding the true worst-case violations and thus the tightenings are not robust to all possible mismatches for which the distributed optimization algorithm could terminate. Hence, the distributed OPF solution could violate constraints even after applying the bound tightening algorithm. However, choosing $\beta < 1$ allows the tightened bounds to be less conservative. This may lead to more optimal distributed OPF solutions. In addition, when the fully robust ($\beta = 1$) bound tightening algorithm leads to infeasibility of the resulting AC OPF problem, an appropriately selected uncertainty budget allows for less conservative bound tightening and may result in feasible AC OPF problems. Our empirical results in the following section indicate that $\beta$ can be made fairly small in practice without introducing significant constraint violations. Thus, we can use larger convergence tolerances to substantially reduce the number of distributed OPF iterations, while achieving negligible constraint violations and only minor suboptimality compared to the OPF problem without tightened bounds. 

\section{Numerical Results}\label{sec:results}

We use Julia with the optimization modeling package JuMP~\cite{DunningHuchetteLubin2017} to formulate the worst-case violation optimization problems. We run distributed OPF to evaluate violations on the bound-tightened test cases using PowerModelsADA~\cite{alkhraijah_harris_coffrin_molzahn-pmada}, and we run centralized OPF problems using PowerModels \cite{8442948}. Our test cases are the 14-bus, 118-bus and 500-bus network models selected from the PGLib-OPF archive~\cite{ieee_pes_pglib-opf_task_force_power_2019}. We divide the 14-bus and 118-bus cases into 3 regions and the 500-bus case into 8 regions for distributed optimization. 

We first examine the relationship between convergence tolerance $\epsilon^{pri}$ and optimality of the bound-tightened cases. To do so, we sweep $\epsilon^{pri}$ across $[5\times10^{-4},5\times10^{-2}]$ for case14 and case118 and across $[10^{-6},10^{-4}]$ for case500. We choose smaller values of $\epsilon^{pri}$ for case500 because violations begin to appear with smaller $\epsilon^{pri}$ for this case. We run the bound tightening algorithm for each value of $\epsilon^{pri}$ and solve a centralized AC OPF problem on the bound-tightened test case with nominal loads. 
Figure 6 shows the cost percent difference for bound-tightened cases compared to original cases across multiple budgets~$\beta$. We compute the cost percent difference as 
$ (\tilde{f}-f^*)/{f^*}$, 
where $\tilde{f}$ is the AC OPF objective value for the bound-tightened case and $f^*$ is the objective value for the original case.
When $\epsilon^{pri}$ is sufficiently large, the bounds are tightened until the resulting test case is not feasible. We mark tolerances that result in infeasible test cases with $\times$. 

\begin{figure}
     \centering
     \begin{subfigure}[b]{\linewidth 
     }
         \centering
         \includegraphics[height=5cm]{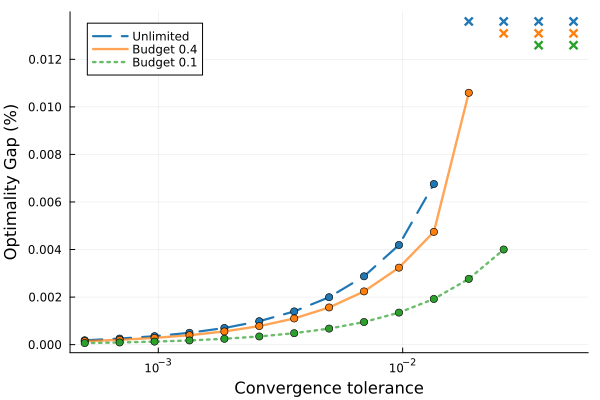}
         \vspace{-0.4em}
         \caption{case14}
         \vspace{0.2em}
         \label{fig:cost_v_tau_14}
     \end{subfigure}
     \hfill
     \begin{subfigure}[b]{\linewidth}
         \centering
         \includegraphics[height=5cm]{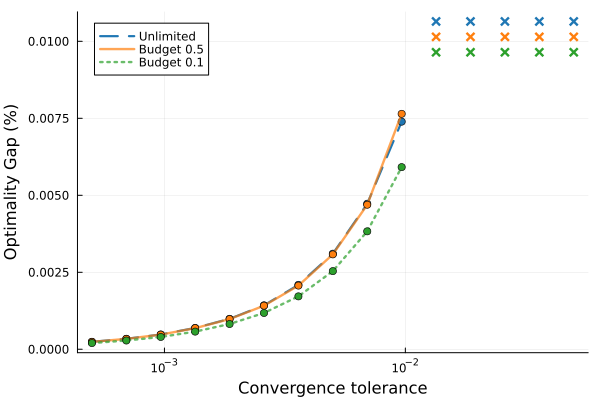}
        \vspace{-0.4em}
         \caption{case118}
         \vspace{0.2em}
         \label{fig:cost_v_tau_118}
     \end{subfigure}
     \hfill
     \begin{subfigure}[b]{\linewidth}
         \centering
         \includegraphics[height=5cm]{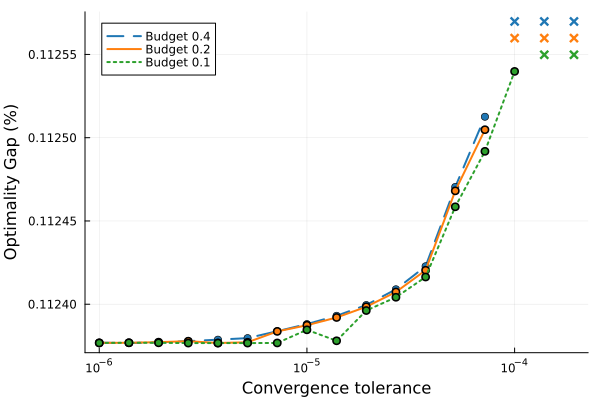}
        \vspace{-0.4em}
         \caption{case500}
         \label{fig:cost_v_tau_500}
     \end{subfigure}
        \caption{Cost vs. convergence tolerance $\epsilon^{pri}$}
        \label{fig:cost_v_tol}
        \vspace{-2em}
\end{figure}

As expected, for every test case, the amount of bound tightening increases with $\epsilon^{pri}$, worsening the suboptimality of the solution. However, the bound-tightened cases' costs are no more than 0.2\% above optimal for all $\epsilon^{pri}$ at which the bounds can be tightened without causing AC OPF infeasibility. Decreasing the budget parameter allows for less conservative bound tightening, which may improve optimality very slightly (by less than 0.05\% for our test cases). More significantly, using a smaller budget can sometimes result in feasible tightened cases for values of $\epsilon^{pri}$ at which tightening with a larger budget or no budget causes infeasibility; see, e.g., convergence tolerance values greater than $10^{-2}$ for the 14-bus case in Figure~\ref{fig:cost_v_tau_14}.

There is some unexpected behavior in these results: for case118 at  $\epsilon^{pri} = 9.7\times 10^{-3}$, the cost is slightly higher for a budget of $\beta = 0.5$ compared to an unlimited budget. 
This is because the bounds are tightened less for the unlimited budget due to an instance in which the solver for the unlimited budget found a local solution, rather than the true global optimum, to one of the optimization problems used to compute the bound tightenings. 
As described in Section~\ref{sec:discussion}, since we use the non-convex AC power flow equations in our optimization formulation, we cannot guarantee that the solver will find the global solution to these worst-case violation problems.

In addition to evaluating the optimality of bound-tightened cases, we also assess whether the bound tightening algorithm prevents constraint violations once the distributed OPF solution is applied to the system. We expect distributed OPF on cases tightened without any mismatch budget ($\beta = 1$) to have no constraint violations. As mentioned above, there is one caveat: due to the non-convex nature of the AC power flow equations in the worst-case violation problems, the solver may find local solutions. However, we find that in practice the bound tightening algorithm using the AC power flow formulation does not result in constraint violations once the distributed OPF solution is applied to the system. We do expect that when the mismatch budget becomes small enough, the constraints will not be tightened sufficiently. Thus, we may see that distributed OPF on test cases tightened with very small mismatch budgets result in constraint violations on the system. 

To assess this, we run distributed OPF computations on cases tightened across a range of values for $\epsilon^{pri}$ and across a range of budgets. Each time we run distributed OPF, we vary the loads by up to 50\% from nominal for case14 and case118 and by up to 30\% from nominal for case500. The perturbation for each load is randomly selected from a uniform distribution across this range. Once the distributed OPF converges to a tolerance of $\epsilon^{pri}$, we solve an AC power flow using the generator dispatch from the distributed OPF solution. We record the average percent violation of any constraints violated and the total number of violations as described in Section~\ref{sec:select_tol}. The results are shown in Table~\ref{table:3}. For every test case, bound tightening with very small budgets ($\beta < 0.05$) may result in a few small violations, but tightening with a budget of at least $\beta = 0.1$ achieves negligible constraint violations. 


\begin{table}[t] 
	\centering
	\small
	\caption{Violations vs. budget}
    \label{table:1}
	    \begin{tabular}{|c | c | c | c |}
     \hline 
    \begin{tabular}{@{}c@{}} \textbf{Test Case} \\ \textbf{(Tolerance)} \end{tabular} & \begin{tabular}{@{}c@{}} \textbf{Budget}\\ $\boldsymbol{\beta}$ \end{tabular} & \begin{tabular}{@{}c@{}} \textbf{Median Number} \\ \textbf{of Violations} \end{tabular}  & \begin{tabular}{@{}c@{}} \textbf{Median Percent} \\ \textbf{Violation} \end{tabular} \\
    \hline 
    \multirow{4}{*}{\begin{tabular}{@{}c@{}} case14 \\ ($10^{-2}$) \end{tabular}} & 0.01 & 1 & 0.484\phantom{0}\\
    & 0.03 & 1 & 0.016\phantom{0} \\
    & 0.06 & 1 & 0.0006 \\
    & 0.10 & 0 & 0\phantom{.0000}\\
    \hline 
    \hline 
    \multirow{4}{*}{\begin{tabular}{@{}c@{}} case118 \\ ($10^{-2}$) \end{tabular}} & 0.01 & 3 & 0.487\phantom{0}  \\
    & 0.03 & 1 & 0.029\phantom{0} \\
    & 0.06 & 0 & 0\phantom{.0000} \\
    & 0.10 & 0 & 0\phantom{.0000} \\
    \hline 
    \hline 
    \multirow{4}{*}{\begin{tabular}{@{}c@{}} case500 \\ ($10^{-4}$) \end{tabular}} & 0.01 & 6 & 1.04\phantom{00}  \\
    & 0.03 & 1 & 0.085\phantom{0} \\
    & 0.06 & 0 & 0.0001 \\
    & 0.10 & 0 & 0\phantom{.0000} \\
    \hline 
        \end{tabular}
\end{table}


We note that one motivation for bound tightening is to reduce the number of iterations to convergence. Bound tightening allows us to increase $\epsilon^{pri}$ without risking constraint violations once the distributed OPF solution is applied to the grid. We showed an example of the impacts of increasing $\epsilon^{pri}$ in Section~\ref{sec:select_tol} for case500. Here, we show in Table~\ref{table:2} the median percent reduction in iterations to convergence when we increase $\epsilon^{pri}$ to the maximum value at which we can feasibly tighten bounds. Just as in Section~\ref{sec:select_tol}, we run distributed OPF repeatedly, perturbing the loads each time by up to 50\% for case14 and case118 and by up to 30\% for case500. For each case, we find $\epsilon^{pri}_{orig}$, the greatest value of $\epsilon^{pri}$ which results in no violations under distributed OPF for the original test case, which is $5 \times 10^{-4}$ for case14 and case118, and $5 \times 10^{-6}$ for case500. Then, we find $\epsilon^{pri}_{tight}$, the greatest value of $\epsilon^{pri}$ for which we can feasibly run a bound tightening algorithm with a budget of 10\% ($\beta = 0.1$) or higher, which is $10^{-2}$ for case14 and case118 and $10^{-4}$ for case500. We measure the percent reduction as 
$(\nu_{\epsilon^{pri}_{orig}}-\nu_{\epsilon^{pri}_{tight}})/\nu_{\epsilon^{pri}_{orig}}$, 
where $\nu_{\epsilon^{pri}}$ is the median number of iterations required to converge to a tolerance of $\epsilon^{pri}$ in our experiments. That is, the percent reduction in iterations in Table~\ref{table:2} indicates the amount by which bound tightening allows us decrease the number of iterations (by increasing $\epsilon^{pri}$) without resulting in constraint violations. For these representative test cases, bound tightening can reduce the number of iterations by over 35\% without increasing the cost by more than 0.2\%. 

\begin{table}[t] 
	\centering
	\small
	\caption{Reduction in Iterations}
    \label{table:2}
	    \begin{tabular}{|c | c | c | c |}
	    \hline
    \textbf{Test case} & case14 & case118 & case500 \\ 
	    \hline 
    \begin{tabular}{@{}c@{}} \textbf{Reduction} \\ \textbf{in Iterations} \end{tabular} & 53.9\% & 85.2\% & 36.9\% \\
    \hline 
        \end{tabular}
        \vspace{-1em}
\end{table}

We also provide a brief discussion on computation time. While collecting these results, we ran the bound tightening algorithm on the test cases for many different values of $\epsilon^{pri}$ and for several different budgets. We record in Table~\ref{table:3} the minimum, median and maximum times required to run the bound tightening algorithm on each test case. We ran the experiments on Georgia Tech’s PACE cluster, where each node had a 16-core 2.7~GHz processor and 64 GB RAM. Recall that all bound tightening occurs offline. To speed up offline bound tightening, we parallelize the computation of worst-case bound violations and adaptively determine which bounds are at risk for violations to reduce the number of problems to be solved. For example, any bound not violated for the original test case will not be violated after other bounds are tightened and can thus be ignored after the first iteration of the bound tightening. 

During real-time operation, when running distributed OPF on a bound-tightened test case, the computation time for solving ADMM subproblems at each iteration is no different from the computation time for subproblems on the original test case. However, a bound-tightened test case allows for selecting a larger convergence tolerance, resulting in fewer iterations required to converge, without risking constraint violations.  

\begin{table}[h] 
	\centering
	\small
	\caption{Bound Tightening Time in Minutes}
    \label{table:3}
	    \begin{tabular}{|c | c | c | c |}
	    \hline
    \textbf{Test case} & \textbf{Minimum} & \textbf{Median} & \textbf{Maximum} \\ 
	    \hline 
	case14 & 0.17 & 0.18 & 0.20 \\ 
        \hline 
        case118 & 1.75 & 2.25 & 2.49 \\
        \hline 
        case500 & 66.2\phantom{00} & 120.8\phantom{000} & 229.7\phantom{000} \\ 
        \hline 
        \end{tabular}
\end{table}

\section{Conclusion}\label{sec:conclusion}
Distributed optimization algorithms provide several advantages, including scalability, flexibility, and privacy, for operating power systems with widespread distributed energy resources. Such algorithms require separate computing agents to reach consensus, up to some convergence tolerance, on the values of shared boundary variable values. Increasing the convergence tolerance generally reduces the number of iterations to convergence, which is a key challenge for distributed algorithms, but may also lead to constraint violations with respect to the original problem. In this paper, we first formulate an optimization problem which finds the worst-case constraint violations that result from applying a distributed OPF solution converged to a given tolerance to the power system. Next, we propose a bound tightening algorithm which, provided that global solutions are found for worst-case violation problems, guarantees that the distributed OPF solution will not cause constraint violations on the real power system. We also introduce a "budget uncertainty" method to bound cumulative boundary variable mismatches in the worst-case violation problem, allowing for less conservative bound tightening. Our numerical results demonstrate that the bound tightening algorithm increases suboptimality only slightly, while allowing for a significant reduction in distributed OPF iterations without causing constraint violations. 

For sufficiently large convergence tolerances, the algorithm tightens bounds to the point that OPF is no longer feasible. Our future work is to increase the range of convergence tolerances for which the bound tightening algorithm maintains OPF feasibility. To do so, we plan to analyze distributions of boundary variable mismatches, explore chance-constrained variants of the worst-case violation problems, and leverage the optimality of solutions to regions' OPF subproblems, which may yield less conservative worst-case violations. 

\printbibliography

\end{document}

%% file: defs.tex
\newcommand{\vct}[1]{\boldsymbol{#1}}
\newcommand{\mtx}[1]{\boldsymbol{#1}}

\newcommand{\set}[1]{\mathcal{#1}}

\newcommand{\vc}{\vct{c}}

\newcommand{\vp}{\vct{p}}
\newcommand{\vq}{\vct{q}}

\newcommand{\vx}{\vct{x}}
\newcommand{\vy}{\vct{y}}
\newcommand{\vz}{\vct{z}}

\newcommand{\vlambda}{\vct{\lambda}}

\newcommand{\vzero}{\vct{0}}

\newcommand{\mA}{\mtx{A}}
\newcommand{\mB}{\mtx{B}}

\newcommand{\setA}{\set{A}}

\newcommand{\setC}{\set{C}}

\newcommand{\setE}{\set{E}}

\newcommand{\setI}{\set{I}}

\newcommand{\setM}{\set{M}}
\newcommand{\setN}{\set{N}}

\newcommand{\setP}{\set{P}}

\newcommand{\setS}{\set{S}}